\documentclass[aps]{revtex4}
\usepackage{amssymb}
\usepackage{color}
\usepackage[colorlinks=true,linkcolor=red]{hyperref}
\usepackage{hyperref}
\begin{document}
\title{Gravitational Trapping of the Bulk Fields in 6D}
\author{Pavle  Midodashvili}
\email{midodashvili@hotmail.com} \affiliation{Department of
Physics,
  Tskhinvali State University,
  2 Besiki Str., Gori
   383500, GEORGIA}
\date{\today}
\begin{abstract}
We present two new solutions to Einstein's equations in
$(1+5)$-spacetime with a positive bulk cosmological constant. One
solution has increasing and another solution decreasing bounded
scale function $\phi(r)$ without singularities in the range from
the origin $r=0$ to  the radial infinity $r=+\infty$. For the both
solutions it is shown that all local fields are localized near the
origin r=0 in the extra space through the gravitational
interaction.
\end{abstract} \maketitle

The idea that our observable 4-dimensional universe may be a brane
extended in some higher-dimensional spacetime with non-compact
extra dimensions and non-factorizable geometry has a long history
\cite{RSh-Akama-Visser-RS} and has been the subject of many recent
studies.
 Large extra dimensions offer an opportunity
for a new solutions to old problems (smallness of cosmological
constant, the origin of the hierarchy problem, the nature of
flavor, etc.). In such theories  it is assumed that all the matter
fields are constrained to live on the $3$-brane and the
corrections to four dimensional Newton's gravity low from bulk
gravitons are small for macroscopic scales. But this models still
need some natural mechanism of localization of known particles on
the brane. The brane solutions and questions of matter
localization on the branes in higher-dimensional bulk spacetimes
have been investigated in various papers
\cite{Oda-ADD-GRSh-GMSh-Giovani-R_DSh-GSh-CP-CK-BG-Pomarol-Greg-CW-Gogber,Paul1-Paul2,Paul3,Paul4}.
 In our opinion the localization mechanism must
be universal for all types of 4-dimensional matter fields. In our
world the gravity is known to be the unique interaction which has
universal coupling with all matter fields. So, if extended extra
dimensions exist, it is natural to assume that trapping of matter
on the brane has a gravitational nature. It is of great interest
that recently in \cite{GogberSingleton} it was found the model in
the brane world where all the local bulk fields (ranging from the
spin 0 scalar field to the spin 2 gravitational field) are
localized on the 3-brane only by the the gravity. The solution was
found in (1+5)-spacetime for the positive bulk cosmological
constant $\Lambda > 0$ and has increasing scale factor $\phi (r)$
asymptotically approaching finite value at the radial infinity. In
the paper \cite{Oda} this solution was extended to the case of a
general (p-1)-brane model with codimension $n$ in general $D=p+n$
space-time dimensions.

In our previous article \cite{Paul4} we have introduced the new
3-brane solution in (1+5)-spacetime which also localizes all the
local bulk fields but in contrast to the solution found in
\cite{GogberSingleton} has decreasing scale factor $\phi (r)$
asymptotically approaching finite non-zero value at the radial
infinity. Such solution exists for the negative cosmological
constant $\Lambda < 0$. In this article we present two new regular
solutions (one with increasing and another with decreasing scale
factor $\phi (r)$) to Einstein's equations in $(1+5)$-spacetime
with a positive bulk cosmological constant $\Lambda > 0$, and for
the both solutions we explicitly show that whole local fields
(spin $0$ scalar field, spin ${\raise0.7ex\hbox{$1$}
\!\mathord{\left/
 {\vphantom {1 2}}\right.\kern-\nulldelimiterspace}
\!\lower0.7ex\hbox{$2$}}$ spinor field, spin 1 gauge field, spin
${\raise0.7ex\hbox{$1$} \!\mathord{\left/
 {\vphantom {3 2}}\right.\kern-\nulldelimiterspace}
\!\lower0.7ex\hbox{$2$}}$ gravitino field and spin $2$
gravitational field), as well as totally antisymmetric tensor
fields, are localized near the origin r=0 in the extra space. In
addition we investigate the technical reasons of the localization
of all bulk fields.

 Let us begin with the details of our solution. In 6D the Einstein
equations with a bulk cosmological constant $\Lambda$ and
stress-energy tensor $T_{AB}$  \begin{equation}
\label{6Dequations}R_{AB}  - \frac{1}{2}g_{AB} R = \frac{1}{{M^4
}}(\Lambda g_{AB}  + T_{AB} )\end{equation} can be derived from
the whole action of the gravitating system
\begin{equation}S = \int {d^6 x\sqrt { - g} \left[ {\frac{{M^4 }}{2}\left( {R +
2\Lambda } \right) + L} \right]}\end{equation}  $R_{AB}$ , $R$,
$M$ and $L$ are respectively the Ricci tensor, the scalar
curvature, the fundamental scale and the Lagrangian of  matter
fields (including brane). All of these physical quantities refer
to $(1+5)$- space with signature $(+ - ... -)$, capital Latin
indices run over $A,B,...=0,1,2,3,5,6$. Suppose that the equations
(\ref{6Dequations})  admit a solution that is consistent with
four-dimensional Poincar\'{e} invariance. Introducing for the
extra dimensions the polar coordinates $(r,\theta )$, where $0 \le
r <  + \infty$ , $0 \le \theta  < 2\pi$ , the six-dimensional
metric satisfying this ansatz we can choose in the form
\cite{Paul1-Paul2,GogberSingleton} :
\begin{equation}\label{ansatzA}ds^2  = \phi ^2 \left( r \right)\eta _{\alpha \beta } \left(
{x^\nu  } \right)dx^\alpha  dx^\beta   - g\left( r \right)\left(
{dr^2  + r^2 d\theta ^2 } \right) ,\end{equation} where small
Greek indices $\alpha ,\beta ,... = 0,1,2,3$ numerate coordinates
and physical quantities in four-dimensional space, the functions
$\phi (r )$ and $g(r)$ depend only on  $r$  and are cylindrically
symmetric in the extra-space, the metric signature is $(+ - ...
-)$.The function $g(r)$ must be positive to fix the signature of
the metric (\ref{ansatzA}).

The source of the brane  is described by a stress-energy tensor
$T_{AB}$ also cylindrically symmetric in the extra-space. Its
nonzero components we choose in the form
\begin{equation}\label{branetensor}T_{\alpha \beta }  =  - g_{\alpha \beta } F_0 \left( r \right),\ \
T_{ij}  =  - g_{ij} F\left( r \right) , \end{equation} where we
have introduced two source functions $F_0$  and $F$, which depend
only on the radial coordinate $r$.

By using cylindrically symmetric metric ansatz (\ref{ansatzA})
 and stress-energy tensor (\ref{branetensor}),  the Einstein equations become
\begin{equation}\label{4dpartA}3\frac{{\phi ''}}{\phi } + 3\frac{{\phi '^2 }}{{\phi ^2 }} +
3\frac{{\phi '}}{{r\phi }} + \frac{{g''}}{{2g}} - \frac{{g'^2
}}{{2g^2 }} + \frac{{g'}}{{2rg}} = \frac{g}{{M^4 }}\left( {F_0  -
\Lambda } \right) + \frac{g}{{\phi ^2 }}\frac{{\Lambda _{phys}
}}{{M_P^2 }} ,\end{equation}

\begin{equation}\label{55partA}6\frac{{\phi '^2 }}{{\phi ^2 }} + 2\frac{{g'\phi '}}{{g\phi }} +
4\frac{{\phi '}}{{r\phi }} = \frac{g}{{M^4 }}\left( {F - \Lambda }
\right) + 2\frac{g}{{\phi ^2 }}\frac{{\Lambda _{phys} }}{{M_P^2 }}
,\end{equation}

\begin{equation}\label{66partA}4\frac{{\phi ''}}{\phi } + 6\frac{{\phi '^2 }}{{\phi ^2 }} -
2\frac{{g'\phi '}}{{g\phi }} = \frac{g}{{M^4 }}\left( {F - \Lambda
} \right) + 2\frac{g}{{\phi ^2 }}\frac{{\Lambda _{phys} }}{{M_P^2
}} ,\end{equation} where the prime denotes differentiation $d/dr$.
The constant $\Lambda _{phys}$ represents the physical
four-dimensional cosmological constant, where
\begin{equation}\label{4deinstequationsA}R_{\alpha \beta }^{(4)}  -
\frac{1}{2}g_{\alpha \beta } R^{(4)}  = \frac{{\Lambda _{phys}
}}{{M_P^2 }}g_{\alpha \beta }.\end{equation} In this equation
$R_{\alpha \beta }^{(4)}$, $R^{(4)}$ and $M_P$ are
four-dimensional physical quantities: Ricci tensor, scalar
curvature and Planck scale.

In the case $\Lambda _{phys}=0$  from the equations
(\ref{4dpartA}), (\ref{55partA}) and (\ref{66partA}) we can find
\begin{equation}\label{sfconA} F' + 4\frac{{\phi '}}{\phi }\left( {F - F_0 } \right) = 0,
\end{equation}
\begin{equation}\label{cfconA}g = \frac{{\delta \phi '}}{r} , \    \ \delta=const,
\end{equation}\begin{equation} \label{equationA}r\frac{{\phi ''}}{\phi } + 3r\frac{{\phi '^2 }}{{\phi ^2 }} +
\frac{{\phi '}}{\phi } = \frac{{rg}}{{2M^4 }}\left( {F - \Lambda }
\right) ,\end{equation} where $\delta$ denotes the integration
constant. The (\ref{sfconA}) represents the connection between
source functions, it is simply a consequence of the conservation
of the stress-energy tensor and can be also independently derived
directly from $D_A T_B^A = 0$.

Let us take the cylindrically symmetric source functions in the
form
\begin{equation}\label{Source1}F\left( r \right) = f_1 \phi ^{ - 1}  - f_2 \phi ^{ -
2},
\end{equation}
\begin{equation}\label{Source2}F_0 \left( r \right) = \frac{3}{4}f_1 \phi ^{ - 1}
- \frac{1}{2}f_2 \phi ^{ - 2},\end{equation}where $f_1$ and $f_2$
are some positive constants. These source functions satisfy the
equation (\ref{sfconA}).

Taking the first integral of the last equation \cite{Paul1-Paul2},
we get
\begin{equation}\label{equationAA}r\phi ' =  - \frac{{\delta \Lambda }}{{10M^4 }}\left( {\phi ^2  -
\frac{{5f_1 }}{{4\Lambda }}\phi  + \frac{{5f_2 }}{{3\Lambda }}}
\right) + \frac{C}{{\phi ^3 }} ,
\end{equation} where  $C$ is the integration constant. Then the equation (\ref{equationAA}) can be presented in the form
\begin{equation}\label{equationAA1}
r\phi ' =  - \frac{{\delta \Lambda }}{{10M^4 }}\left( {\phi  - d_1
} \right)\left( {\phi  - d_2 } \right) + \frac{C}{{\phi ^3 }},
\end{equation}
where $d_1$ and $d_2$ denote the roots of the quadratic equation
${\phi ^2 - \frac{{5f_1 }}{{4\Lambda }}\phi  + \frac{{5f_2
}}{{3\Lambda }}}=0$  with respect to the variable $\phi$:
\begin{equation}\label{d1 and d2}
d_1  = \frac{{5f_1 }}{{8\Lambda }}\left( {1 - \sqrt {1 -
\frac{{64\Lambda f_2 }}{{15 f_1^2 }}} } \right),\ \ d_2  =
\frac{{5f_1 }}{{8\Lambda }}\left( {1 + \sqrt {1 - \frac{{64\Lambda
f_2 }}{{15 f_1^2 }}} } \right).
\end{equation} Let us suppose $\Lambda > 0$  and  $\frac{{64\Lambda f_2 }}{{15f_1 }}  < 1$. Then for the $d_1$ and $d_2$  we have  $d_2
> d_1 >0$.

Setting $C=0$,  introducing the parameters
\begin{equation}\label{Parameters}
a= \frac{{\delta \Lambda }}{{10M^4 }}, \ \ b=a(d_2-d_1),
\end{equation}
and imposing the boundary conditions for the solution in the form
\begin{equation}
\left. {\phi \left( r \right)} \right|_{r = 0}  = const > 0,\ \
\left. {\phi \left( r \right)} \right|_{r =  + \infty }  = const
> 0 ,
\end{equation}
\begin{equation}
\left. {g(r)} \right|_{r = 0}  = const,\ \ \left. {g(r)}
\right|_{r = + \infty }  = const,
\end{equation}
we can easily find  two solutions of the equation
(\ref{equationAA}) in the following cases:

\ \ \ \ \ \ \ \ \ \ \ \ \ \ \ \ \ \ \ i) $\Lambda  > 0$, \ \ $ b
\ge 2 $, \ \ $\delta
> 0$
\begin{equation}\label{Solution}\phi \left( r \right) = \frac{{d_1  + d_2 cr^b }}{{1 + cr^b
}},\ \ \ g\left( r \right) = \delta b \left( {d_2  - d_1 }
\right)\frac{{cr^{b - 2} }}{{\left( {1 + cr^b } \right)^2 }};
\end{equation}

\ \ \ \ \ \ \ \ \ \ \ \ \ \ \ \ \ \ \ ii) $\Lambda  > 0$, \ \ $ b
\le  - 2 $, \ \ $\delta < 0$
\begin{equation}\label{Solution1}
\phi \left( r \right) = \frac{{d_1 r^{\left| b \right|}  + d_2
c}}{{r^{\left| b \right|}  + c}},\ \ g\left( r \right) = \left|
\delta \right|\left| b \right|\left( {d_2  - d_1 }
\right)\frac{{cr^{\left| b \right| - 2} }}{{\left( {r^{\left| b
\right|} + c } \right)^2 }};
\end{equation} where
$c>0$  is some positive integration constant. As we can see these
solutions exists in the case of positive bulk cosmological
constant $\Lambda >0 $. The scale factors $\phi(r)$ and $g(r)$ of
these solutions are smooth bounded functions of radial coordinate
$r$. The scale factor $\phi (r)$ of the first solution
monotonically increases from the value equal to $d_1 >0$ at the
origin $r=0$ in the extra space and asymptotically approaches the
value equal to $d_2 >0$ at the radial infinity $r=+\infty$, while
the scale factor $\phi(r)$ of the second solution monotonically
decreases from the value equal to $d_2>0$ at the origin $r=0$ in
the extra space and asymptotically approaches the value equal to
$d_1>0$ at the radial infinity $r=+\infty$. As we mention below
one of the reasons of the  localization of all local fields is
connected with the properties of the scale factor $g(r)$. Let us
briefly consider the properties of this function for the first
solution (\ref{Solution}) (for the second solution
(\ref{Solution1}) the results are the same) for various values of
parameter $b$. In the case $|b|=2$, at the origin $r=0$ and at the
infinity $r=+\infty$ in the extra space this function has values
equal to $2c\delta (d_2-d_1)$ and $0$ respectively. The derivative
of this function $ g'(r)$ is negative in the range $0 < r < +
\infty$ except at the origin of extra space $r=0$ where it is
equal to $0$. So in this case $g(r)$ is monotonically decreasing
function between $r=0$ and $r=+\infty$. In the case $|b|>2$, this
function is equal to $0$ at the origin $r=0$ in the extra space,
it increases in the range
 $0 <r < r_{\max }$,
approaches the maximum value at the $r_{max}$, then decreases in
the range $r_{max} < r < +\infty$ and monotonically approaches the
asymptotic value equal to $0$ at the radial infinity. Here we have
introduced $r_{max}$ which can be found from the equation
$g'(r)=0$ and is given by
\begin{equation}\label{r_max}
r_{\max }  = \left( {\frac{{b - 2}}{{b + 2}}}
\right)^{\frac{1}{b}} \left( {\frac{1}{c}} \right)^{\frac{1}{b}}.
\end{equation}
In this case the derivative $ g'(r)$ is smooth bounded function
except for the case $|b|<3$  when the derivative $ g'(r)$
approaches the asymptotic value equal to infinity only at the
origin $r=0$.

From the formula (\ref{r_max})it is easy to notice, that in the
case of fixed parameter $b$ the $r_{max}$ tends to the origin
$r=0$ in the extra space when the positive integration constant
$c$ tends to infinity. In the case $c\gg1$ the scale factor $g(r)$
has a $\delta$-function-like behavior and therefore it can provide
the sharp localization of bulk fields near the origin $r=0$ in the
extra space (for the second solution (\ref{Solution1}) we have the
same results in the case $c \ll 1 $).

Now we turn our attention to the problem of the localization of
the bulk fields on the brane in the background geometries
(\ref{Solution}) and (\ref{Solution1}). Of course, in due
analysis, we will neglect the back-reaction on the geometry
induced by the existence of the bulk fields, and from now on,
without loss of generality, we shall take a flat metric on the
brane. In the following we examine only the first solution
(\ref{Solution}), since the results for the second solution are
the same (\ref{Solution1}).

We start with a massless, spin $0$, real scalar coupled to
gravity:
\begin{equation}\label{RealScalar}S_0   =  - \frac{1}{2}\int {d^6 x\sqrt { - {}^6g} g^{AB}
\partial _A \Phi \partial _B \Phi }.\end{equation} The corresponding equation of
motion has the form
\begin{equation}\label{RealScalarEqMotion}\frac{1}{{\sqrt { - {}^6g} }}\partial _M \left( {\sqrt { - {}^6g}
g^{MN} \partial _N \Phi } \right) = 0.\end{equation}  It turns out
that in the background metric (\ref{Solution}) the zero-mode
solution of (\ref{RealScalarEqMotion}) is $\Phi _0 \left( {x^M }
\right) = \upsilon \left( {x^\mu  } \right)\rho _0$, where $\rho
_0=const$,  and $\upsilon \left( {x^\mu  } \right)$ satisfies the
Klein-Gordon equation on the brane $\eta ^{\mu \nu }
\partial _\mu  \partial _\nu  \upsilon \left( {x^\mu  } \right) =
0$. Substituting this solution into the starting action
(\ref{RealScalar}), the action can be cast to
\begin{equation}\label{RealScalarZeroModeAction}S_0  =  - \frac{1}{2}\rho _0^2 \int_0^{2\pi } {d\theta }
\int_0 ^{ + \infty } {dr\phi ^2 gr} \int {d^4 x\sqrt { - \eta }
\eta ^{\mu \nu } \partial _\mu  \upsilon \left( {x^\alpha }
\right)\partial _\nu  \upsilon \left( {x^\alpha  } \right)}  + ...
=\end{equation}
\begin{equation}\label{RealScalarZeroModeAction1}=  - \pi \delta\rho _0^2 \int_{\phi \left(
0 \right)}^{\phi \left( { + \infty } \right)} {\phi ^2 d\phi }
\int {d^4 x\sqrt { - \eta } \eta ^
 {\mu \nu } \partial _\mu  \upsilon \left( {x^\alpha  }
\right)\partial _\nu  \upsilon \left( {x^\alpha  } \right)}  +
...= \end{equation}
\begin{equation}\label{RealScalarZeroModeAction1Second} =  - \frac{{\pi \delta \rho _0^2
}}{3}\left( {d_2^3 - d_1^3 } \right)\int {d^4 x\sqrt { - \eta }
\eta ^{\mu \nu } \partial _\mu  \upsilon \left( {x^\alpha  }
\right)\partial _\nu  \upsilon \left( {x^\alpha  } \right)}  +
...\ \ .\end{equation} The integral over $r$ in
(\ref{RealScalarZeroModeAction}) is finite, so the 4-dimensional
scalar field is localized at the origin $r=0$ in the extra space.

For spin ${\raise0.7ex\hbox{$1$} \!\mathord{\left/
 {\vphantom {1 2}}\right.\kern-\nulldelimiterspace}
\!\lower0.7ex\hbox{$2$}}$ fermion starting action is the Dirac
action given by
\begin{equation}\label{FermionAction}S_{\frac{1}{2}}  = \int {d^6 x\sqrt { - {}^6g} \overline \Psi
i\Gamma ^M D_M \Psi } ,\end{equation} from which the equation of
motion is given by
\begin{equation}\label{FermionEqMotion}\Gamma ^M D_M \Psi  = \left( {\Gamma ^\mu  D_\mu   + \Gamma ^r D_r
+ \Gamma ^\theta  D_\theta  } \right)\Psi  = 0 \ \ .\end{equation}
We introduce the vielbein $h_A^{\widetilde A}$ through the usual
definition $g_{AB}  = h_A^{\widetilde A} h_B^{\widetilde B} \eta
_{\widetilde A\widetilde B}$ where $\widetilde A,\widetilde B,...$
denote the local Lorentz indices. $\Gamma ^A$  in a curved
space-time is related to $\gamma ^A$ by $\Gamma ^A  =
h_{\widetilde A}^A \gamma ^{\widetilde A}$. The spin connection
$\omega _M^{\widetilde M\widetilde N}$ in the covariant derivative
\ \  $D_M \Psi  = \left( {\partial _M  + \frac{1}{4}\omega
_M^{\widetilde M\widetilde N} \gamma _{\widetilde M\widetilde N} }
\right)\Psi$ \ \ is defined as
$$\omega _M^{\widetilde M\widetilde N}  = \frac{1}{2}h^{N\widetilde M}
\left( {\partial _M h_N^{\widetilde N}  - \partial _N
h_M^{\widetilde N} } \right) - \frac{1}{2}h^{N\widetilde N} \left(
{\partial _M h_N^{\widetilde M}  - \partial _N h_M^{\widetilde M}
} \right) - \frac{1}{2}h^{P\widetilde M} h^{Q\widetilde N} \left(
{\partial _P h_{Q\widetilde R}  - \partial _Q h_{P\widetilde R} }
\right)h_M^{\widetilde R}.$$ After these conventions are set we
can decompose the $6$-dimensional spinor into the form $\Psi
\left( {x^M } \right) = \psi \left( {x^\mu  } \right)A\left( r
\right)\sum {e^{il\theta } } $.  We require that the
four-dimensional part satisfies the massless equation of motion
$\gamma ^\mu  \partial _\mu  \psi \left( {x^\beta  } \right) = 0$
and the chiral condition $\gamma ^r \psi \left( {x^\mu  } \right)
= \psi \left( {x^\mu  } \right)$. As a result we obtain the
following equation for the $s$-wave mode
\begin{equation}\label{FermionZeroModeEq}\left[ {\partial _r  + 2\frac{{\phi '}}{\phi } +
\frac{1}{2}\frac{{\partial _r \left( {rg^{\frac{1}{2}} }
\right)}}{{rg^{\frac{1}{2}} }}} \right]A\left( r \right) = 0.
\end{equation} The solution to this equation reads:
\begin{equation}\label{FermionZeroModeEqSolution}A\left( r \right) = A_0 \phi ^{ - 2} g^{ - \frac{1}{4}} r^{ -
\frac{1}{2}},\end{equation} with $A_0$ being an integration
constant. Substituting this solution into the Dirac action
(\ref{FermionAction}) we have
\begin{equation}\label{FermionActionZeroMode}S_{\frac{1}{2}}  = 2\pi A_0^2 \int_0 ^{ + \infty }
{dr\phi ^{ - 1} g^{\frac{1}{2}} } \int {d^4 x\sqrt { - \eta }
\overline \psi  i\gamma ^\mu  \partial _\mu  \psi }  + ...
=\end{equation}
\begin{equation}\label{FermionActionZeroModeSecond}
 = 2\pi A_0^2 \sqrt {\delta \left( {d_2  - d_1 } \right)c}
 \int_0^{ + \infty } {\frac{{r^{\frac{1}{2}b} dr}}{{r\left( {d_1
 + d_2 cr^b } \right)}}} \int {d^4 x\sqrt { - \eta } \overline \psi  i\gamma ^\mu
  \partial _\mu  \psi }  + ... \ \ .\end{equation}
 In the case $b > 2 $  the integral over $r$ is obviously finite. Indeed, the
integrand in (\ref{FermionActionZeroModeSecond}) scales as $r^{ -
\frac{1}{2}b - 1}$ at the radial infinity and is the smooth
functions between $r= 0 $ and $r =  + \infty $, so this integral
over $r$ is finite. Hence the spin ${\raise0.7ex\hbox{$1$}
\!\mathord{\left/
 {\vphantom {1 2}}\right.\kern-\nulldelimiterspace}
\!\lower0.7ex\hbox{$2$}}$  fermion is localized at the origin
$r=0$ in the extra space only by the gravitational interaction.

Now let us consider  the action of $U(1)$  vector field:
\begin{equation}\label{VectorFieldAction}S_1  =  - \frac{1}{4}\int {d^6 x\sqrt { - {}^6g} g^{AB} g^{MN}
F_{AM} F_{BN} } ,\end{equation} where $F_{MN}  = \partial _M A_N -
\partial _N A_M$  as usual. From this action the equation of
motion is given by
\begin{equation}\label{VectorFieldEqMotion}\frac{1}{{\sqrt { - {}^6g} }}\partial _M \left( {\sqrt { - {}^6g}
g^{MN} g^{RS} F_{NS} } \right) = 0.\end{equation} By choosing the
gauge condition $A_{\theta}=0$ and decomposing the vector field as
\begin{equation}\label{decomposition1}A_\mu  \left( {x^M } \right) = a_\mu  \left( {x^\mu  }
\right)\sum\limits_{l,m} {\sigma _m \left( r \right)e^{il\theta }
},\end{equation}
 \begin{equation}\label{decomposition2}A_r \left( {x^M } \right) = a_r \left( {x^\mu  }
\right)\sum\limits_{l,m} {\sigma _m \left( r \right)e^{il\theta }
},\end{equation} it is straightforward to see that there is the
$s$-wave $(l=0)$ constant  solution $\sigma _m \left( r
\right)=\sigma _0=const$ and  $a_r=const$.  In deriving this
solution we have used $\partial _\mu  a^\mu   = \partial ^\mu
f_{\mu \nu }  = 0$ with the definition of $f_{\mu \nu }  =
\partial _\mu  a_\nu   - \partial _\nu  a_\mu$. As in the previous
cases, let us substitute this constant solution into the action
(\ref{VectorFieldAction}). It turns out that the action is reduced
to
\begin{equation}\label{VectorFieldActionZeroMode}S_1  =  - \frac{\pi }{2}\sigma _0^2 \int_0 ^{ + \infty }
{drgr} \int {d^4 x\sqrt { - \eta } \eta ^{\alpha \beta } \eta
^{\mu \nu } f_{\alpha \mu } f_{\beta \nu } }+... =\end{equation}
\begin{equation}\label{VectorFieldActionZeroModeSecond}= -\frac{{\pi \delta
}}{2}\sigma _0^2 \int_{\phi  \left( 0 \right)}^{\phi \left( { +
\infty } \right)} {d\phi  } \int {d^4 x\sqrt { - \eta } \eta
^{\alpha \beta } \eta ^{\mu \nu } f_{\alpha \mu } f_{\beta \nu } }
+ ... =  - \frac{{\pi  \delta }}{2}\sigma _0^2 \left( {d_2 - d_1}
\right)\int {d^4 x\sqrt { - \eta } \eta ^{\alpha \beta } \eta
^{\mu \nu } f_{\alpha \mu } f_{\beta \nu } } + ... \ \
.\end{equation}

 As we can see
from (\ref{VectorFieldActionZeroModeSecond}) the integral over $r$
in (\ref{VectorFieldActionZeroMode}) is finite. Thus, the vector
field is also localized.

Next we consider spin ${\raise0.7ex\hbox{$3$} \!\mathord{\left/
 {\vphantom {3 2}}\right.\kern-\nulldelimiterspace}
\!\lower0.7ex\hbox{$2$}}$ field (the gravitino). We begin with the
action of the Rarita-Schwinger gravitino
\begin{equation}\label{GravitinoAction}S_{\frac{3}{2}}  = \int {d^6 x\sqrt { - {}^6g} \overline \Psi  _A
i\Gamma ^{\left[ A \right.} \Gamma ^B \Gamma ^{\left. C \right]}
D_B \Psi _C },\end{equation}from which the equation of motion is
given by
\begin{equation}\label{GravitinoEqMotion}\Gamma ^{\left[ A \right.} \Gamma ^B \Gamma ^{\left. C \right]}
D_B \Psi _C  = 0.\end{equation} Here the square bracket denotes
the anti-symmetrization and the covariant derivative is defined
with the affine connection $\Gamma _{BC}^A  = h_{\widetilde B}^A
\left( {\partial _B h_C^{\widetilde B}  + \omega _B^{\widetilde
B\widetilde C} h_{C\widetilde C} } \right)$  by
\begin{equation}\label{Derivative}D_A \Psi _B  = \partial _A \Psi _B  - \Gamma _{AB}^C \Psi _C  +
\frac{1}{4}\omega _A^{\widetilde A\widetilde B} \gamma
_{\widetilde A\widetilde B} \Psi _B\end{equation} After taking the
gauge condition $\Psi _\theta   = 0$ we look for the solutions of
the form  $\Psi _\mu  \left( {x^A } \right) = \psi _\mu  \left(
{x^\nu  } \right)u\left( r \right)\sum {e^{il\theta } }$, $\Psi _r
\left( {x^A } \right) = \psi _r \left( {x^\nu  } \right)u\left( r
\right)\sum {e^{il\theta } }$ where  $\psi _\mu  \left( {x^\nu  }
\right)$  satisfies the following equations $\gamma ^\nu  \psi
_\nu   =
\partial ^\mu  \psi _\mu   = \gamma ^{\left[ \nu  \right.} \gamma
^\rho  \gamma ^{\left. \tau \right]}
\partial _\rho  \psi _\tau   = 0$, $\gamma ^r \psi _\nu   = \psi
_\nu$. For the $s$-wave solution and $\psi _r \left( {x^\nu  }
\right)=0$ we have for the equation of motion
(\ref{GravitinoEqMotion}) the following form
\begin{equation}\label{GravitinoEqMotionZeroMode}\left[ {\partial _r  + \frac{3}{2}\frac{{\phi '}}{\phi } +
\frac{1}{2}\frac{{\partial _r \left( {rg^{\frac{1}{2}} }
\right)}}{{rg^{\frac{1}{2}} }}} \right]u\left( r \right) =
0.\end{equation} The solution to this equation is
\begin{equation}\label{GravitinoZeroMode}u\left( r \right) = u_0 \phi ^{ - \frac{3}{2}} g^{ - \frac{1}{4}}
r^{ - \frac{1}{2}},\end{equation} with $u_0$  being an integration
constant. Substituting this solution into the action
(\ref{GravitinoAction}) we get
\begin{equation}\label{GravitinoActionZeroMode}S_{\frac{3}{2}}  = 2\pi u_0^2 \int_0^{ + \infty }
{dr\phi ^{ - 2} g^{\frac{1}{2}} } \int {d^4 x\sqrt { - \eta }
\overline \psi  _\mu  i\gamma ^{\left[ \mu  \right.} \gamma ^\nu
\gamma ^{\left. \rho  \right]} \partial _\nu  \psi _\rho  }  +
...=\end{equation}
\begin{equation}\label{GravitinoActionZeroModeSecond}
 = 2\pi u_0^2 \sqrt {\delta \left( {d_2  - d_1 } \right)c} \int_0^{ + \infty }
 {dr\frac{{r^{\frac{1}{2}b} \left( {1 + cr^b } \right)}}{{r\left( {d_1  + d_2 cr^b }
 \right)^2 }}} \int {d^4 x\sqrt { - \eta } \overline \psi  _\mu  i\gamma ^{\left[ \mu
 \right.} \gamma ^\nu  \gamma ^{\left. \rho  \right]} \partial _\nu  \psi _\rho  }  + ...
 \ \ .\end{equation}
 As in the case for the spin
${\raise0.7ex\hbox{$1$} \!\mathord{\left/
 {\vphantom {1 2}}\right.\kern-\nulldelimiterspace}
\!\lower0.7ex\hbox{$2$}}$ fermion
(\ref{FermionActionZeroModeSecond}), when $b > 2$ the integrand in
(\ref{GravitinoActionZeroModeSecond}) scales as $ r^{ -
\frac{1}{2}b - 1} $ at the radial infinity and is smooth function
between $r= 0 $ and $ r = + \infty$, so the integral over $r$ is
finite. This means that spin ${\raise0.7ex\hbox{$3$}
\!\mathord{\left/
 {\vphantom {3 2}}\right.\kern-\nulldelimiterspace}
\!\lower0.7ex\hbox{$2$}}$ field is also localized at the origin $r
= 0 $ in the extra space.

Now let us consider spin $2$ gravitational field. In this case we
consider the spin-$2$ metric fluctuations $H_{\mu \nu }$:
\begin{equation}\label{MetricFluctuation}ds^2  = \left\{ {\phi ^2 \left( r \right)\eta _{\alpha \beta }
\left( {x^\nu  } \right) + H_{\alpha \beta } } \right\}dx^\alpha
dx^\beta   - g\left( r \right)\left( {dr^2  + r^2 d\theta ^2 }
\right).\end{equation} The corresponding equation of motion for
the fluctuations has the following form:
\begin{equation}\label{MetricFluctuationEqMotion}\frac{1}{{\sqrt { - {}^6g} }}\partial _A \left( {\sqrt { - {}^6g}
g^{AB} \partial _B H_{\mu \nu } } \right) = 0.\end{equation} From
the 4-dimensional point of view fluctuations are described by a
tensor field $H_{\mu \nu }$ which is transverse and traceless:
\begin{equation}\partial _\mu  H_\nu ^\mu   = 0,\ \ H_\mu ^\mu   =
0.
\end{equation}
This tensor is invariant under 4-dimensional general coordinate
transformations. We look for solutions of the form
\begin{equation} \label{MetricFluctuation1}H_{\alpha \beta } =
h_{\alpha \beta } (x^\nu )\sum\limits_{ml} {\tau _m (r)\exp
(i\theta l)},
\end{equation} where $\partial ^2 h_{\mu \nu } (x^\alpha  ) = m_0^2 h_{\mu \nu } (x^\alpha
)$. It is easy to show that the equation of motion
(\ref{MetricFluctuationEqMotion}) has the zero-mass ($m_0=0$) and
$s$-wave ($l=0$) constant solution $\tau_0=const$. Substitution of
this zero mode into the Einstein-Hilbert action leads to
\begin{equation}\label{EinsteinHilbertAction}S_2  \sim 2\pi \tau _0^2 \int_0 ^{ + \infty } {dr\phi ^2
gr} \int {d^4 x\left[ {\partial ^\rho  h^{\alpha \beta } \partial
_\rho  h_{\alpha \beta }  + ...} \right]}=\end{equation}
\begin{equation}\label{EinsteinHilbertAction1}
 = 2\pi \delta \tau _0^2 \int_{\phi \left( 0 \right)}^{\phi \left( { +
 \infty } \right)} {\phi ^2 d\phi } \int {d^4 x\left[ {\partial ^\rho
 h^{\alpha \beta } \partial _\rho  h_{\alpha \beta }  + ...} \right]}
 = \frac{2}{3}\pi \delta \tau _0^2 \left( {d_2^3  - d_1^3 } \right)\int
 {d^4 x\left[ {\partial ^\rho  h^{\alpha \beta } \partial _\rho  h_{\alpha \beta }  + ...} \right]}
.\end{equation} The integral over $r$ is finite, so the bulk
graviton is localized at the origin $r=0$ in the extra space. With
our background metric the general expression for the
four-dimensional Planck scale $M_P$, expressed in terms of $M$, is
\begin{equation}\label{PlanckScale}M_P^2  = 2\pi M^4 \int_ 0 ^{ + \infty } {dr\phi ^2 gr}  =
2\pi \delta M^4 \int_ 0 ^{ + \infty } {dr\phi ^2 \phi '} =
\frac{2}{3} \pi \delta M^4 (d_2^3 - d_1^3).\end{equation}  The
inequality $ M \ll M_P $ is possible by adjusting $ M^4 $ and the
product $ \delta (d_2^3 - d_1^3 ) $, and thus could lead to a
solution of the gauge hierarchy problem.

Finally, let us consider the totally antisymmetric tensor fields.
The action of $k$-rank totally antisymmetric tensor field $A_k$ is
of the form
\begin{equation}\label{TotallyAntisymmetricFieldAction}S_k  =  -
\frac{1}{2}\int {F_{k + 1}  \wedge  * F_{k + 1} ,}
\end{equation} where $ F_{k + 1}  = dA_k $.
The corresponding equation of motion is given by
\begin{equation}\label{TotallyAntisymmetricFielEqMotion} d \wedge  * F_{k + 1}
 = 0 . \end{equation} It is easy to show that  $A_{\mu _1 \mu _2 ...\mu _k }  = a_{\mu _1 \mu _2 ...\mu _k }
\left( {x^\nu  } \right)u_0$  with $u_0=const$ is a solution to
the equation of motion (\ref{TotallyAntisymmetricFielEqMotion}) if
$ d \wedge * f = 0$ where $ f = da$. Substituting this solution in
the action (\ref{TotallyAntisymmetricFieldAction}) leads to the
expression
\begin{equation}\label{TotallyAntisymmetricFieldActionZeroMode1}
S_k  \sim \int_ 0 ^{ + \infty } {dr\phi ^{2 - 2k} gr} \int {f_{k +
1}  \wedge  * f_{k + 1}  + ...}  =
\end{equation}
\begin{equation}\label{TotallyAntisymmetricFieldActionZeroMode2}
 = \delta \int_{\phi \left( 0  \right)}^{\phi \left( { +
 \infty } \right)} {\phi ^{2 - 2k} d\phi } \int {f_{k + 1}  \wedge  * f_{k + 1}  + ...}
 = \frac{{ \delta  }}{{\left( {3 - 2k} \right)}}\left( {d_2^{3-2k} - d_1^{3 - 2k} }
 \right)\int {f_{k + 1}  \wedge  * f_{k + 1}  + ...}
\end{equation} As we can see from
(\ref{TotallyAntisymmetricFieldActionZeroMode2}) the integral over
$r$ in (\ref{TotallyAntisymmetricFieldActionZeroMode1}) is finite,
so the totally antisymmetric tensor fields are also localized by
the gravitational interaction.

In conclusion, in this article we have presented two new solutions
to the Einstein's equations in the $(1+5)$ -spacetime with
increasing and decreasing scale factors $\phi \left( r \right)$.
These solutions are found for the positive bulk cosmological
constant $\Lambda  > 0$. In addition for the our solutions we have
presented a complete analysis of localization of a bulk fields at
the origin in the extra space via only the gravitational
interaction. There are two technical reasons of the localization
of all bulk fields. The first reason lies in the fact that the
scale factors $\phi \left( r \right)$  are  smooth bounded
functions without singularities from the $r=0$ to the radial
infinity. The second one, which in our opinion is the main reason,
is connected with the function $g(r)$. As it was mentioned above
in the case $c \gg 1$ for the first solution (\ref{Solution}) and
in the case $c \ll 1$ for the second solution
(\ref{Solution1})this function has a $\delta$-function-like
behavior, and in contrast to the model considered in \cite{Oda} in
our model the wave functions of the zero-mode solutions of the
bulk fields along the extra dimensions are peaked at the location
of the brane.

{\bf Acknowledgements:} Author would like to acknowledge the
hospitality extended during his visits at the Abdus Salam
International Centre for Theoretical Physics where the main part
of this work was done.

\end{document}